\documentclass[pra,aps,nopacs,onecolumn,twoside,superscriptaddress]{article}

\usepackage{amsmath}
\usepackage{hyperref} % 添加超链接支持
\usepackage{cleveref}
\usepackage{authblk}
\usepackage{multirow}
\usepackage{amsfonts,amssymb,caption,color,epsfig,graphics,graphicx,hyperref,latexsym,mathrsfs,url,verbatim,epstopdf,enumerate,amsthm}
\usepackage{hyperref,color,epsfig,verbatim}
\usepackage{fontenc}
\usepackage{xcolor}% 字体颜色宏包
\usepackage{braket}
\usepackage{bm}% 加粗部分公式
\usepackage{graphicx,float}

\usepackage{tabularx} % 用于自动调整列宽

\usepackage{multirow}
\usepackage{booktabs}
\usepackage{epstopdf}
\usepackage{epsfig}
\usepackage{longtable}% 长表格
\usepackage{supertabular}% 跨页表格
\usepackage{tikz}%画CNOT的
\usepackage{subfigure}
\usepackage{algorithmic}
\usepackage{changepage}% 换页
\usepackage{enumerate}% 短编号
\usepackage{caption}% 设置标题
\usepackage{tikz}
\hypersetup{colorlinks,linkcolor={blue},citecolor={blue},urlcolor={red}}

\usepackage[left=2.50cm,right=2.50cm,top=2.80cm,bottom=2.50cm]{geometry}% 页边距设置
\usepackage{fancyhdr} %设置全文页眉、页脚的格式
\newtheorem{definition}{Definition}
\newtheorem{proposition}[definition]{Proposition}
\newtheorem{lemma}[definition]{Lemma}

\newtheorem{theorem}[definition]{Theorem}

\newtheorem{remark}[definition]{Remark}
\newtheorem{example}[definition]{Example}

\def\Dbar{\leavevmode\lower.6ex\hbox to 0pt
{\hskip-.23ex\accent"16\hss}D}
\makeatletter
\def\url@leostyle{%
  \@ifundefined{selectfont}{\def\UrlFont{\sf}}{\def\UrlFont{\small\ttfamily}}}
\makeatother
\def\Dbar{\leavevmode\lower.6ex\hbox to 0pt
{\hskip-.23ex\accent"16\hss}D}

\begin{document}

\title{Realization of permutation groups by quantum circuit}% 题目
\author[1]{Junchi Liu \thanks{junchiliu@buaa.edu.cn}}
\author{Yangyang Ren \thanks{renyangyang@buaa.edu.cn}}
\author[1]{Yan Cao  \thanks{yancao@buaa.edu.cn}}
\author[1]{Hanyi Sun \thanks{21091021@buaa.edu.cn}}
\author[1,2]{Lin Chen \thanks{linchen@buaa.edu.cn (corresponding author)}}
\affil[1]{LMIB and School of Mathematical Sciences, Beihang University, Beijing 100191, China, Beihang University, Beijing 100191, China}
\affil[2]{International Research Institute for Multidisciplinary Science, Beihang University, Beijing 100191, China}
\date{}

\maketitle
%\pacs{03.65.Ud, 03.67.Mn}

\begin{abstract}
The permutation group is of wide interest and application in many science branches. 
We show that six CNOT gates in a quantum circuit without single-qubit gates are both necessary and sufficient for the implementation of permutation group of three elements. Then we provide all circuits realizing the element by using exactly six CNOT gates. We also extend our result to the realization of quantum circuits for the permutation group of any elements. 
\end{abstract}

\lhead{}% 页眉左边设为空
\chead{}% 页眉中间设为空
\rhead{}% 页眉右边设为空
\lfoot{}% 页脚左边设为空
\cfoot{\thepage}% 页脚中间显示页码
\rfoot{}% 页脚右边设为空
	
\Large

keywords: quantum circuit, permutation group, CNOT gate

%\tableofcontents
 
 \section{Introduction}
	
	%\subsection{Introduction to the problem}

	%\subsection{Current research situation}

	%\subsection{Research method}

Quantum circuit has been used to realize various tasks such as quantum-information masking \cite{shi2021k}, teleportation and dense coding of both states and operations \cite{ren2017ground,yeo2006teleportation,luo2019qt,Schaetz2004quantum,mattle1996dense,fiaschi2021optomechanical,huelga2001quantum,bennett1992communication,pra2022,bennett1993teleporting}. Such tasks rely on the implementation of quantum operations such as single-qubit unitray gate, two-qubit controlled-NOT(CNOT) gate, and swap gate \cite{cy15,rrg07,zlc00,dw13,cybenko2001reducing}. In particular, the swap gate can be regarded as an element of order-two symmetric group from group theory. The theory has been one of the fundamental branches in abstract algebra for centuries, with plenty of applications to the science such as physics, chemistry, electrical engineering, and so on. Hence, one naturally expects that the
symmetric group finds applications in a larger scope of life and science, when the practical cost of such a group is necessarily considered. For example in quantum information theory, the quantum cost of an arbitrary gate was first proposed by Barenco et al \cite{barenco1995elementary}.
One can simply regard the quantum cost as the number of unitary gates used in designing a quantum circuit. Hence, it is expected to have a lower quantum cost in practice. In particular, researchers are interested in finding the optimal quantum circuit which requires the least quantum cost \cite{2011Quantum}.
It has been shown that the two-qubit swap gate can be realized by exactly three CNOT gates both necessarily and sufficiently \cite{Farrokh2004Optimal,vd04}. It is therefore an interesting problem to ask whether one can construct elements of general symmetric groups by spending a small number of CNOT gates.  

A multi-qubit system comprises multiple qubits, each of which is capable of existing in superposition states, thereby creating an expanded quantum state space.  The evolution of multi-qubit systems is facilitated by quantum gate operations, enabling information transfer between qubits and the establishment of quantum entanglement.  Control and manipulation of multi-qubit systems allow for various applications, including quantum computing, communication, and simulation. Through the extension to multi-qubit system, we can categorize complex situations into several equivalence classes. Within each equivalence class, the number of CNOT gates required for performing circuit swaps and renaming operations remains consistent.

In this paper, we exclusively utilize CNOT gates for implementing permutation groups generated by more than two elements. In Lemma  \autoref{le:2qubitswap}, we recall that three CNOT gates are both necessary and sufficient to execute a two-qubit swap gate operation. Subsequently, in Lemma  \autoref{le:C(n)le3(n-1)}, we show that the maximum number of CNOT gates needed to carry out an $n$-qubit substitution operation is $3(n-1)$. Moving forward, our analysis in \Cref{sec:three} reveals that utilizing five or fewer CNOT gates is insufficient for implementing a three-qubit swap gate corresponding to the permutation element $(123)$. Hence six CNOT gates are both necessary and sufficient for implementing  $(123)$. This is done by employing a graph-theoretic approach to rigorously validate the results in terms of at most five CNOT gates. Using computational tools, we exhaustively explore all valid circuit diagrams containing exactly six CNOT gates to successfully execute the swap gate for $(123)$, by explaining the equivalence classes in Remark \autoref{rmk:3qubit} and \Cref{tab:my_table}. We conclude them in Theorem \autoref{thm:3qubit}. To extend our analysis to the multiqubit scenario, we present the reducible and irreducible permutation elements in Definition  \autoref{df:multiqubit}. We clarify the equivalence between rows in the multi-qubit space and provide an approximate upper bound for multi-qubits to perform the aforementioned operations in Theorem \autoref{thm:nqubit}. The comprehensive exploration of this paper aims to pave the way for further advancements in understanding quantum circuit optimization via multiple use of a specific two-qubit gate.

%In this paper, we realize the elements of permutation group of more than two elements by using only the CNOT gates. We show that in the three-qubit case, employing six CNOT gates is both necessary and sufficient. In other word, five or fewer CNOT gates are not enough for realizing the three-qubit swap gate. This critical analysis, focusing on configurations involving 1-5 CNOT gates individually, is conducted using graph theory methods to rigorously validate the findings.  Utilizing computational tools, we  exhaust all admissible circuit diagrams comprising six CNOT gates to successfully execute the operation in three qubits. For extending our analysis to the multiqubit scenarios, we elucidate the equivalence between lines in multiqubit spaces. 

The rest of this paper is organized as follows. In \Cref{sec:pre}
we introduce the basic notions and facts used in this paper. 
In \Cref{sec:three} we show the implementation of three-qubit swap gate using CNOT gates. Then we extend our implementation to the multiqubit scenario in
\Cref{sec:highdim}. Finally we conclude in \Cref{sec:con}.

\section{Preliminaries}
    \label{sec:pre}

\begin{figure}[H]
		\begin{center}
			\begin{tikzpicture}[scale=2]
				
				\foreach \y in {1,2} %几条线就几个数字，此处是3维情况，则0,1,2;  4维则0,1,2,3;
				{
					\draw (0,\y) -- (2,\y);%每条线的长度，此处长度为0-10，可适当修改，\y表示对于每条线都同样
					\draw (1,0.8) -- (1,2);%表示（1,1）（1,2）两点之间连线，注意坐标原点位于整幅图的左下角，但图中未做显示
					\fill (1,2) circle(2pt);
					\draw (1,1) circle [radius=0.2cm];%做空心圆的操作
					%\filldraw[fill=black] (2,1) circle [radius=0.5cm];%实心圆	
				}		
				
			\end{tikzpicture}
		\end{center}
		\caption{Line representation of the CNOT gate}
            \label{fig:1}
	\end{figure}
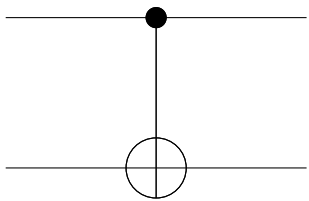

In this section, we introduce the main notions and facts for this paper. It is known that every $n$-qubit pure state is in the $n$-partite Hilbert space $(C^{2})^{\otimes n}$ spanned by the orthonormal basis $\{e_{j_1}\otimes\ldots     \otimes e_{j_n},j_1,\ldots,j_n= 0, 1\}$, and
		$e_{0}:=\ket{0}={\begin{pmatrix}
				1\\
				0
		\end{pmatrix}}$, $e_{1}:=\ket{1}={\begin{pmatrix}
				0\\
				1
		\end{pmatrix}}$. To produce an entangled state, an interaction on the multiqubit state is usually required. A typical interaction is the two-qubit CNOT gate in  \Cref{fig:1}, which is capable of generating one ebit of entanglement by acting on a product state. 
For $x,y=0,1$, the effect of CNOT gate is $CNOT(\ket{x}\ket{y}) = \ket{x}\ket{x\oplus y}$.
Here the computational basis vectors are represented by $\ket{x}\ket{y}:=\ket{x,y}:=\ket{xy}$ such that
	\begin{align}
		\ket{00}= \left[ \begin{array}{ccc}
			1 \\
			0 \\
			0 \\
			0\end{array} \right],\quad
		\ket{01}= \left[ \begin{array}{ccc}
			0 \\
			1 \\
			0 \\
			0\end{array} \right], \\
		\ket{10}= \left[ \begin{array}{ccc}
			0 \\
			0 \\
			1 \\
			0\end{array} \right],\quad
		\ket{11}= \left[ \begin{array}{ccc}
			0 \\
			0 \\
			0 \\
			1\end{array} \right]. 
	\end{align}

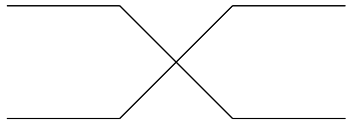
\begin{figure}[H]
		\begin{center}
			\begin{tikzpicture}[scale=1.5]
				
				\foreach \y in {0,1} %几条线就几个数字，此处是3维情况，则0,1,2;  4维则0,1,2,3;
				{   
					\draw (0,0) -- (1,0);
					\draw (0,1) -- (1,1);
					\draw (2,0) -- (3,0);
					\draw (2,1) -- (3,1);
					\draw (1,1) -- (2,0);
					\draw (1,0) -- (2,1);%表示（1,1）（1,2）两点之间连线，注意坐标原点位于整幅图的左下角，但图中未做显示
					% (1,2) circle(2pt);
					%\draw (1,1) circle [radius=0.2cm];%做空心圆的操作
					%\filldraw[fill=black] (2,1) circle [radius=0.5cm];%实心圆	
				}		
				
			\end{tikzpicture}
		\end{center}
		\caption{Line representation of the SWAP gate}
            \label{fig:2}
	\end{figure}

Next, the SWAP gate in \Cref{fig:2} can exchange qubits and create entanglement. 
It is mathematically written as
	$
		SWAP = \left[\begin{array}{cccc}
			1&0&0&0 \\
			0&0&1&0 \\
			0&1&0&0 \\
			0&0&0&1\end{array}\right].
	$ The following fact is known.
	
	\begin{lemma}
\label{le:2qubitswap} 
The SWAP gate can be implemented by using exactly three CNOT gates to exchange two qubits.
	\end{lemma}

	%\begin{figure}[H]
	%\centering
	%\includegraphics[width=8cm]{S=C}
	%\caption{Three CNOT gates form a SWAP gate}
	%\end{figure}
	\begin{figure}[H]
		\centering
		
		\begin{equation}
			\begin{minipage}[c]{0.2\textwidth}
				\begin{center}
					\begin{tikzpicture}[scale=1.5]
						\foreach \y in {1,2} 
						{
							\centering
							\draw (0,\y) -- (2,\y);
							\draw (0.5,0.8) -- (0.5,2);
							\fill (0.5,2) circle(2pt);
							\draw (0.5,1) circle [radius=0.2cm];
							\draw (1,1) -- (1,2.2);
							\fill (1,1) circle(2pt);
							\draw (1,2) circle [radius=0.2cm];
							\draw (1.5,0.8) -- (1.5,2);
							\fill (1.5,2) circle(2pt);
							\draw (1.5,1) circle [radius=0.2cm];
							%\filldraw[fill=black] (2,1) circle [radius=0.5cm];	
							%\draw (2.2,1.4) -- (2.4,1.4);
							%\draw (2.2,1.6) -- (2.4,1.6);
						}		
					\end{tikzpicture}
				\end{center}
			\end{minipage}
			=
			\begin{minipage}[c]{0.2\textwidth}
				\begin{center}
					\begin{tikzpicture}[scale=1.5]
						
						\foreach \y in {0,1} 
						{   
							\centering
							\draw (0.5,0) -- (1,0);
							\draw (0.5,1) -- (1,1);
							\draw (2,0) -- (2.5,0);
							\draw (2,1) -- (2.5,1);
							\draw (1,1) -- (2,0);
							\draw (1,0) -- (2,1);
							% (1,2) circle(2pt);
							%\draw (1,1) circle [radius=0.2cm];
							%\filldraw[fill=black] (2,1) circle [radius=0.5cm];
						}		
						
					\end{tikzpicture}
				\end{center}
			\end{minipage}
			=
			\begin{minipage}[c]{0.2\textwidth}
				\begin{center}
					\begin{tikzpicture}[scale=1.5]
						\foreach \y in {1,2} 
						{
							\centering
							\draw (0,\y) -- (2,\y);
							\draw (0.5,1) -- (0.5,2.2);
							\fill (0.5,1) circle(2pt);
							\draw (0.5,2) circle [radius=0.2cm];
							\draw (1,0.8) -- (1,2);
							\fill (1,2) circle(2pt);
							\draw (1,1) circle [radius=0.2cm];
							\draw (1.5,1) -- (1.5,2.2);
							\fill (1.5,1) circle(2pt);
							\draw (1.5,2) circle [radius=0.2cm];
							%\filldraw[fill=black] (2,1) circle [radius=0.5cm];
						}		
					\end{tikzpicture}
				\end{center}
			\end{minipage}
			\nonumber
		\end{equation}

		\caption{Three CNOT gates form a SWAP gate}
            \label{fig:3}
	\end{figure}
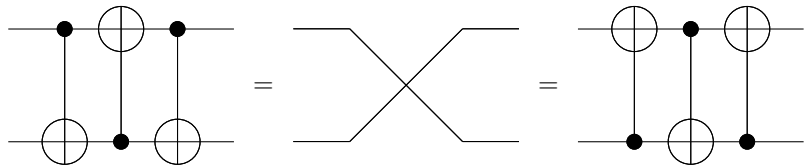
As far as we know, the extension of above lemma is a challenge.
\label{ddd3}
The main target of this paper is to study the least number $C(n)$ of CNOT gates required to realize the multiqubit unitary gate for the element $(12\dots n)$ of the $n$-element permutation group, when no local unitary gates are used. We present a simple observation for the number $C(n)$.
\begin{lemma}
\label{le:C(n)le3(n-1)}
The upper bound of $C(n)$ is $3(n-1)$.
  \end{lemma}
	\begin{proof}
We apply the induction for the claim. The claim for $n=2$ holds by using Lemma \autoref{le:2qubitswap}, indeed, the two-qubit swap gate can be realized by using three CNOT gates. Suppose the claim holds for $n-1$, that is, $C(n-1)\le 3(n-2)$. Because the element $(12...n)=(12...n-1)(n-1,n)$, one can realize it by using $C(n-1)+C(2)=3(n-1)$ CNOT gates. So the claim holds for $n$. We have finished the proof.
 \end{proof}

\begin{figure}[H]
            \centering
	    \begin{tikzpicture}[scale=1.5]
		\foreach \y in {0,1,2} 
		{
			\draw (0,\y) -- (7,\y);
		
			\draw (1,0.8) -- (1,2);
			\fill (1,2) circle(2pt);
			\draw (1,1) circle [radius=0.2cm];
			\node at (1,-0.5){A};
   
			\draw (2,1) -- (2,2.2);
			\fill (2,1) circle(2pt);
			\draw (2,2) circle [radius=0.2cm];
			\node at (2,-0.5){B};
   
			\draw (3,-0.2) -- (3,2);
			\fill (3,2) circle(2pt);
			\draw (3,0) circle [radius=0.2cm];
			\node at (3,-0.5){C};
   
			\draw (4,0) -- (4,2.2);
			\fill (4,0) circle(2pt);
			\draw (4,2) circle [radius=0.2cm];
			\node at (4,-0.5){D};
   
			\draw (5,-0.2) -- (5,1);
			\fill (5,1) circle(2pt);
			\draw (5,0) circle [radius=0.2cm];
			\node at (5,-0.5){E};
   
			\draw (6,0) -- (6,1.2);
			\fill (6,0) circle(2pt);
			\draw (6,1) circle [radius=0.2cm];
			\node at (6,-0.5){F};
		}		
	    \end{tikzpicture}	
            \caption{six CNOT gates in the 3-qubit system}
            \label{fig:six CNOT gates in the 3-qubit system}
	\end{figure}
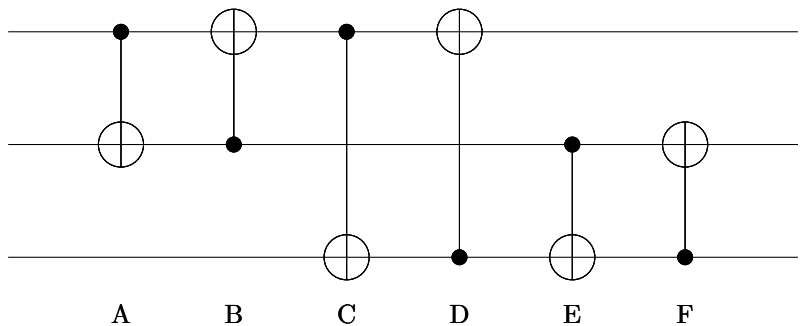

To demonstrate the above lemma, we show some definitions and notations for our next steps. In a three-qubit system, there are only six possible configurations for the CNOT gate. This is because the CNOT gate acts on two qubits, with one qubit serving as the control qubit, and the other qubit serving as the target qubit.
We provide the six configurations in \Cref{tab:Defination of CNOT} and \Cref{fig:six CNOT gates in the 3-qubit system}
 \label{the definition of CNOT}.

\begin{table}[htbp]
    \centering
    \caption{six configurations of CNOT gates}
    \label{tab:Defination of CNOT}
        \resizebox{\textwidth}{!}{%
    \begin{tabular}{ccccccc} % 根据列数调整格式，c表示居中对齐，l表示左对齐，r表示右对齐
        \toprule % 表格顶部横线
        Names of configurations & A & B & C & D & E & F  \\
        \midrule % 中间横线
        the control qubit & first & second & first & third & second & third \\
        the target qubit & second & first & third & first & third & second \\
        \bottomrule % 表格底部横线
    \end{tabular}
    }
    
\end{table}

In an $n$-qubit circuit, some quantum gates may have the same effect by switching the qubits. For example, if the circuit works for a three-qubit swap gate $(123)$, then 
one can show that the CNOT gate on the first and second qubit of circuit is the same as the CNOT gate on the second and third qubit of a new circuit by moving the bottom qubit to the first. More formally, we present the following definition. 
\begin{definition}
     For an $n$-qubit quantum circuit, we can perform an operation where we move the bottom qubit to the first. Each time we perform this operation, we count it as \textbf{one operation}. In an $n$-qubit circuit, by performing $n$ operations, we can obtain $n$ different circuits. We say that they belong to the same \textbf{equivalence class}.
     \label{the definition of equivalence class}
\end{definition}
Evidently, any two quantum circuits from an equivalence class can be transformed into each other through a finite number of operations. It simplifies the classification of various sorts of quantum circuits without eliminating them. To illustrate \Cref{the definition of equivalence class}, 
we show an example. 

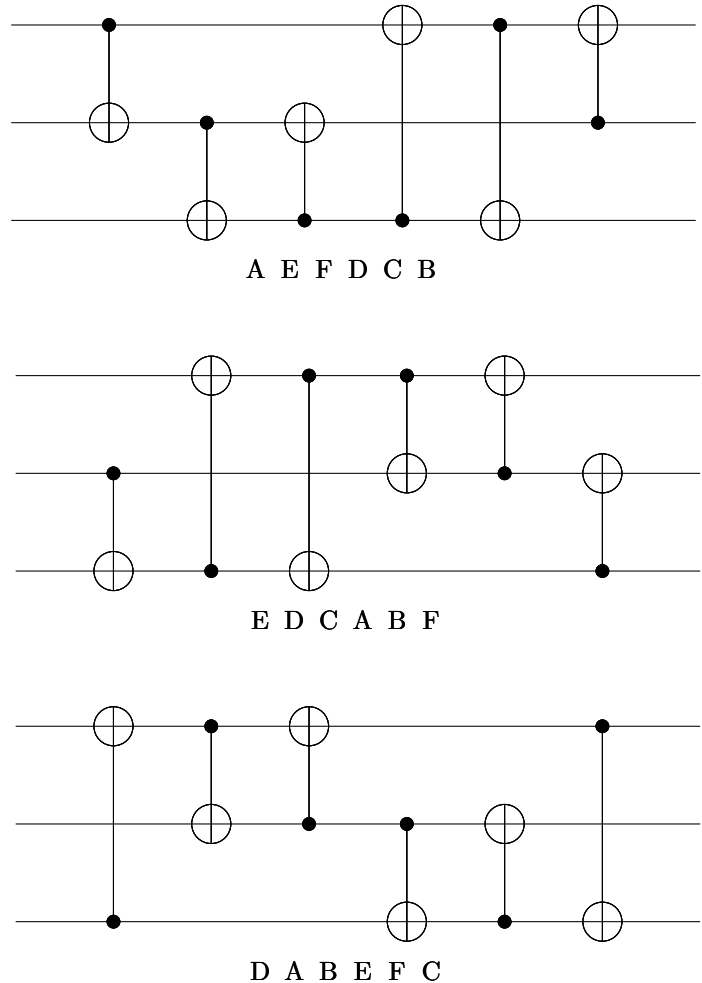
\begin{figure}[htbp]
    \centering
\begin{tikzpicture}[scale=1.3]
		\foreach \y in {0,1,2} 
		{
			\draw (0,\y) -- (7,\y);
		
			\draw (1,0) -- (1,2.2);
			\fill (1,0) circle(2pt);
			\draw (1,2) circle [radius=0.2cm];
			\node at (2.5,-0.5){D};
   
			\draw (2,0.8) -- (2,2);
			\fill (2,2) circle(2pt);
			\draw (2,1) circle [radius=0.2cm];
			\node at (2.85,-0.5){A};
   
			\draw (3,1) -- (3,2.2);
			\fill (3,1) circle(2pt);
			\draw (3,2) circle [radius=0.2cm];
			\node at (3.2,-0.5){B};
   
			\draw (4,-0.2) -- (4,1);
			\fill (4,1) circle(2pt);
			\draw (4,0) circle [radius=0.2cm];
			\node at (3.55,-0.5){E};
   
			\draw (5,0) -- (5,1.2);
			\fill (5,0) circle(2pt);
			\draw (5,1) circle [radius=0.2cm];
			\node at (3.9,-0.5){F};
   
			\draw (6,-0.2) -- (6,2);
			\fill (6,2) circle(2pt);
			\draw (6,0) circle [radius=0.2cm];
			\node at (4.25,-0.5){C};
		}		
	\end{tikzpicture}	
 \captionof{figure}{Three cases of equivalent class 'AEFDCB'}
    \label{fig:equivalent class_example}
\end{figure}

    \begin{example}
        \label{ex:9}
We consider the three-qubit gate consisting of six CNOT gates $AEFDCB$ in the first quantum circuit of \Cref{fig:equivalent class_example}.	One can show that the circuit is equivalent to the remaining two circuits in \Cref{fig:equivalent class_example}.
    \end{example}

\begin{figure}[htbp]
\centering
 \begin{tikzpicture}[scale=1.3]
		\foreach \y in {0,1,2} 
	{
		\draw (0,\y) -- (7,\y);
		
			\draw (1,0.8) -- (1,2);
			\fill (1,2) circle(2pt);
			\draw (1,1) circle [radius=0.2cm];
			\node at (2.5,-0.5){A};
   
			\draw (2,-0.2) -- (2,1);
			\fill (2,1) circle(2pt);
			\draw (2,0) circle [radius=0.2cm];
			\node at (2.85,-0.5){E};
   
			\draw (3,0) -- (3,1.2);
			\fill (3,0) circle(2pt);
			\draw (3,1) circle [radius=0.2cm];
			\node at (3.2,-0.5){F};
   
			\draw (4,0) -- (4,2.2);
			\fill (4,0) circle(2pt);
			\draw (4,2) circle [radius=0.2cm];
			\node at (3.55,-0.5){D};
   
			\draw (5,-0.2) -- (5,2);
			\fill (5,2) circle(2pt);
			\draw (5,0) circle [radius=0.2cm];
			\node at (3.9,-0.5){C};
   
			\draw (6,1) -- (6,2.2);
			\fill (6,1) circle(2pt);
			\draw (6,2) circle [radius=0.2cm];
			\node at (4.25,-0.5){B};
   	}		

	\end{tikzpicture}	
    \label{fig:example of equivalence}
\end{figure}

\begin{figure}[htbp]
\centering
 \begin{tikzpicture}[scale=1.3]
		\foreach \y in {0,1,2} 
	{
   \draw (0,\y) -- (7,\y);
		
			\draw (1,-0.2) -- (1,1);
			\fill (1,1) circle(2pt);
			\draw (1,0) circle [radius=0.2cm];
			\node at (2.5,-0.5){E};

            \draw (2,0) -- (2,2.2);
			\fill (2,0) circle(2pt);
			\draw (2,2) circle [radius=0.2cm];
			\node at (2.85,-0.5){D};
   
			\draw (3,-0.2) -- (3,2);
			\fill (3,2) circle(2pt);
			\draw (3,0) circle [radius=0.2cm];
			\node at (3.2,-0.5){C};
   
			\draw (4,0.8) -- (4,2);
			\fill (4,2) circle(2pt);
			\draw (4,1) circle [radius=0.2cm];
			\node at (3.55,-0.5){A};
   
			\draw (5,1) -- (5,2.2);
			\fill (5,1) circle(2pt);
			\draw (5,2) circle [radius=0.2cm];
			\node at (3.9,-0.5){B};
   
			\draw (6,0) -- (6,1.2);
			\fill (6,0) circle(2pt);
			\draw (6,1) circle [radius=0.2cm];
			\node at (4.25,-0.5){F};
   			
    }
	\end{tikzpicture}	
    
\end{figure}

\section{The implementation of three-qubit swap gate} 
 \label{sec:three}

In this section, we claim that the number $C(3)=6$ above Lemma \autoref{le:C(n)le3(n-1)} for the element $(123)$ in Theorem \autoref{thm:3qubit}. That is, the implementation of three-qubit swap gate requires the cost of exactly six CNOT gates without local unitary gates. Hence, the inequality in Lemma \autoref{le:C(n)le3(n-1)} is saturated for the three-qubit case. 

To prove the claim, it follows from Lemma \autoref{le:C(n)le3(n-1)} that $C(3) \leq 6$. It implies that we only need consider the cases involving one to six CNOT gates. If there is only one CNOT gate in the circuit, then the gate cannot act on all three qubits, and the circuit does not realize the three-qubit swap gate evidently.  
 
In the case of two CNOT gates, there exists one qubit with an uncontrolled target. Note that the three-qubit space is spanned by following eight orthonormal vectors, 
\begin{eqnarray}
\ket{0}\otimes\ket{0}\otimes\ket{0},   
\quad
\ket{0}\otimes\ket{0}\otimes\ket{1},   
\\
\ket{0}\otimes\ket{1}\otimes\ket{0},   
\quad
\ket{0}\otimes\ket{1}\otimes\ket{1},   
\\
\ket{1}\otimes\ket{0}\otimes\ket{0},   
\quad
\ket{1}\otimes\ket{0}\otimes\ket{1},   
\\
\ket{1}\otimes\ket{1}\otimes\ket{0},   
\quad
\ket{1}\otimes\ket{1}\otimes\ket{1}.
\end{eqnarray}
Without loss of generality, we can assume that two CNOT gates take respectively the first and second qubits as the controlled qubits. Thus, the third qubit is the controlling qubit which remains unchanged. Consequently, the circuit does not work as a three-qubit swap gate. 
So two CNOT gates are not enough for realizing the permutation element $(123)$.

In the rest of this section, we show that the three-qubit swap gate cannot be realized by using exactly three, four and five CNOT gates, in Sub\cref{subsec:3cnot}, Sub\cref{sec:4cnot}, and Sub\cref{sec:5cnot}, respectively. Then we implement the three-qubit swap gate by using exactly six CNOT gates in Sub\cref{sec:6cnot}. We demonstrate all circuits for realizing the permutation element $(123)$ in Remark \autoref{rmk:3qubit} and Theorem \autoref{thm:3qubit}.

\subsection{Three CNOT gates}\label{subsec:3cnot}

We recall the $n$-qubit swap gate sending every product vector $\ket{a_1,a_2,...,a_n}$ to $\ket{a_2,...,a_n,a_1}$. Hence, it is sufficient to consider the case $a_j=0,1$ for implementing the gate. Let us consider the case where the input state is 
\begin{equation}
    \ket{1}\otimes...\otimes\ket{1}:=\ket{1,...,1} \label{eq:input=|111>}
\end{equation}
For convenience, we define the function 
\begin{equation}   
\varphi(n,k):=a_1+...+a_n \label{phi} 
\end{equation}
such that $\ket{a_1,...,a_n}$ is an $n$-qubit product state after the input state is performed by exactly $k$ CNOT gates in the $n$-qubit quantum circuit. Note that $a_j=0$ or $1$ for any $j\in[1,n]$.

For example, \Cref{eq:input=|111>} implies that
$\varphi(3,0)=1+1+1=3$. After performing one CNOT gate, we have $\varphi(3,1)=1+1+0=2$, because the CNOT gate converts some $\ket{1,1}$ to $\ket{1,0}$ or $\ket{0,1}$. It follows from Lemma \autoref{le:C(n)le3(n-1)} and \Cref{phi} that $k\in[0,3n-3]$. Besides, the argument of obtaining $\varphi(3,1)$ shows claim (i) of the following theorem, while claim (ii) and (iii) follow from \Cref{eq:input=|111>} and \Cref{phi}.

\begin{figure}[H]
           \begin{center}
		\begin{tikzpicture}
			
			\draw[->] (0,0) -- (5,0) node[right] {i};
			\draw[->] (0,0) -- (0,4) node[above] {$\varphi(3,i)$};

			\draw[blue, fill=blue] (0,3) circle (2pt);
			\draw[blue, fill=blue] (1,2) circle (2pt);
			\draw[blue, fill=blue] (2,2) circle (2pt);
			\draw[blue, fill=blue] (3,3) circle (2pt);

			\draw[red] (0,3) -- (1,2) -- (2,2) -- (3,3);

			\node[below] at (1,0) {1};
			\node[below] at (2,0) {2};
			\node[below] at (3,0) {3};
			\node[left] at (0,1) {1};
			\node[left] at (0,2) {2};
			\node[left] at (0,3) {3};
			
		\end{tikzpicture}
            \caption{Diagram illustrating the change of $\varphi(3,i)$ with respect to i after i-th CNOT gates}
            \label{fig:4}
	   \end{center}
       \end{figure}
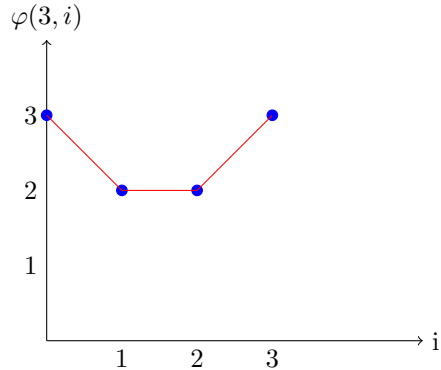

%\red{ There's a problem with the citation format, which seems to be causing a conflict}
\begin{proposition}
\label{pp:|phi(n)-phi(n-1)|<=1}
(i) During the process of applying a CNOT gate, the change in $\varphi(n,k)$ is at most one, i.e. $|\varphi(n,k)-\varphi(n,k+1)|\leq1$. 

(ii) For the $n$-qubit system, we have $0\le\varphi(n,k)\le n$.

(iii) $\varphi(n,C(n)-1)=n-1$.
\end{proposition}
Suppose the quantum circuit for the element $(123)$ consists of exactly three CNOT gates. Recall that the input state is $\ket{1,1,1}$ by \Cref{eq:input=|111>}, and thus the output state is still $\ket{1,1,1}$. By using \Cref{phi}, we have $\varphi(3,3)=3$. It follows from Proposition \autoref{pp:|phi(n)-phi(n-1)|<=1} (i) that $\varphi(3,2)=2$ or $3$. Using Proposition \autoref{pp:|phi(n)-phi(n-1)|<=1} (iii) we obtain that $\varphi(3,2)=2$.
We conclude the above discussion in \Cref{fig:4}.

Due to the equivalence of circuit swapping in \Cref{the definition of equivalence class}, we can assume that the first CNOT gate of circuit acts on the first two qubits. 
By exhausting all cases  satisfying \Cref{fig:4} with the input state as $\ket{1,1,1}$, there is always one qubit that is not controlled. So it is impossible to achieve the permutation process for all cases.

\subsection{Four CNOT gates}
\label{sec:4cnot}
 
\begin{figure}[H]
            \begin{center}
		\begin{tikzpicture}
			% 坐标轴
			\draw[->] (0,0) -- (5,0) node[right] {i};
			\draw[->] (0,0) -- (0,4) node[above] {$\varphi(3,i)$};
			
			% 数据点
			\draw[blue, fill=blue] (0,3) circle (2pt);
			\draw[blue, fill=blue] (1,2) circle (2pt);
                \draw[blue, fill=blue] (2,1) circle (2pt);
			\draw[blue, fill=blue] (3,2) circle (2pt);
			\draw[blue, fill=blue] (4,3) circle (2pt);
			
			% 折线
			\draw[red] (0,3) -- (1,2);
			\draw[red] (3,2) -- (4,3);
                \draw[red] (1,2) -- (2,1);
                \draw[red] (2,1) -- (3,2);
			% 数据点的标签
			\node[below] at (1,0) {1};
   
			\node[below] at (2,0) {2};
			\node[below] at (3,0) {3};
			\node[below] at (4,0) {4};
			\node[left] at (0,1) {1};
			\node[left] at (0,2) {2};
			\node[left] at (0,3) {3};
			
		\end{tikzpicture}
	   \end{center}
            \caption{Diagram illustrating the change of $\varphi(3,i)$ with respect to i after i-th CNOT gates}
            \label{fig:5}
        \end{figure}
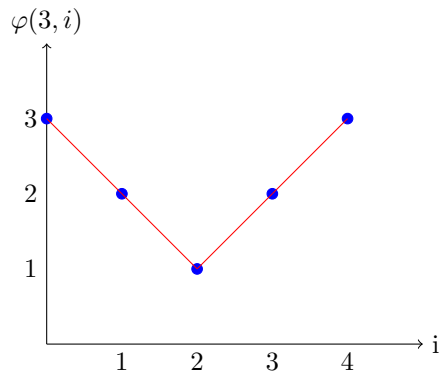

For this scenario, we conduct a discussion on $\varphi(3,k)$ similar to that in \Cref{fig:4}. We present our argument in \Cref{fig:5}, when $\varphi(3,2)=1$ and the input state is still $\ket{1,1,1}$. By exhausting the cases where there are no control bits on certain qubits, it can be observed that there is always at least one qubit remaining as $\ket{1}$. In that case, this qubit will not act as the controlled end, which does not meet our requirements.

For the other two cases $\varphi(3,2)=2$ and $ \varphi(3,2)=3$, we can have a similar discussion and find that in the first scenario, only one qubit can act as the controlled end, while the other two qubits remain unchanged. In the second scenario, two qubits can act as the controlled end, with at least one qubit remaining unchanged. Therefore, neither meets our requirements.

\subsection{Five CNOT gates}
\label{sec:5cnot}

	By applying a similar analysis to $\varphi(3,k)$ and using Proposition \autoref{pp:|phi(n)-phi(n-1)|<=1}, we can determine that it takes three CNOT gates for $\varphi(3,k)$ to transition from $\varphi(3,2)$ to $\varphi(3,4)$. There are two possibilities: either $\varphi(3,k)$ remains unchanged after each operation, or there will be one occurrence each of the changes $+1,-1$ and $0$. By considering the symmetry of the latter case, we can limit our discussion to the following four cases.
 
	According to Proposition \autoref{pp:|phi(n)-phi(n-1)|<=1}(i), we will now  discuss all the possible diagrams that start from $\varphi(3,0)=3$ and end with $\varphi(3,5)=3$. In this situation, $\varphi(3,1)=2$ and $\varphi(3,4)=2$ is settled. Define a sequence $\{\varphi(3,i+1)-\varphi(3,i)\}$ where i=1,2,3. In the next description we briefly denoted the range of the sequence in a row. For example, $(-1)+(+1)+0$ means that $\varphi(3,2)-\varphi(3,1)=-1$, $\varphi(3,3)-\varphi(3,2)=1$ and $\varphi(3,4)-\varphi(3,3)=0$.
 
	$\bullet 0+0+0$
	
        \begin{figure}[H]
            \begin{center}
		\begin{tikzpicture}
			% 坐标轴
			\draw[->] (0,0) -- (5,0) node[right] {i};
			\draw[->] (0,0) -- (0,4) node[above] {$\varphi(3,i)$};
			
			% 数据点
			\draw[blue, fill=blue] (0,3) circle (2pt);
			\draw[blue, fill=blue] (1,2) circle (2pt);
			\draw[blue, fill=blue] (4,2) circle (2pt);
			\draw[blue, fill=blue] (5,3) circle (2pt);
			\draw[blue, fill=blue] (2,2) circle (2pt);
			\draw[blue, fill=blue] (3,2) circle (2pt);
			% 折线
			\draw[red] (0,3) -- (1,2) -- (2,2) -- (3,2) -- (4,2) -- (5,3);
			% 数据点的标签
			\node[below] at (1,0) {1};
			\node[below] at (2,0) {2};
			\node[below] at (3,0) {3};
			\node[below] at (4,0) {4};
			\node[below] at (5,0) {5};
			\node[left] at (0,1) {1};
			\node[left] at (0,2) {2};
			\node[left] at (0,3) {3};
			
		\end{tikzpicture}
	   \end{center}
            \caption{Diagram illustrating the change of $\varphi(3,i)$ with respect to i after i-th CNOT gates}
            \label{fig:6}
        \end{figure}
        
	After discussion, one can obtain that in the case of input qubit $\ket{1}\otimes\ket{1}\otimes\ket{1}$, to obtain the output result $\ket{1}\otimes\ket{1}\otimes\ket{1}$, the control qubit for the 2nd, 3rd, and 4th CNOT gates must be on the circuit where $\ket{0}$ is located, and the fifth CNOT gate must act on either the first two qubits, or the second and third qubits.
	
	When the input quantum information is $\ket{1}\otimes\ket{1}\otimes\ket{0}$, after the fourth CNOT gate, the output quantum information must be $\ket{1}\otimes\ket{0}\otimes\ket{0}$. If we want to obtain the desired permutation result, i.e. $\ket{1}\otimes\ket{0}\otimes\ket{1}$, then the fifth CNOT gate must act on the ac portion. This contradicts with the previous discussion, thus this situation is not valid.
	
	$\bullet(+1)+(-1)+0$
	
        \begin{figure}[H]
            \begin{center}
		\begin{tikzpicture}
			% 坐标轴
			\draw[->] (0,0) -- (5,0) node[right] {i};
			\draw[->] (0,0) -- (0,4) node[above] {$\varphi(3,i)$};
			
			% 数据点
			\draw[blue, fill=blue] (0,3) circle (2pt);
			\draw[blue, fill=blue] (1,2) circle (2pt);
			\draw[blue, fill=blue] (4,2) circle (2pt);
			\draw[blue, fill=blue] (5,3) circle (2pt);
			\draw[blue, fill=blue] (2,3) circle (2pt);
			\draw[blue, fill=blue] (3,2) circle (2pt);
			% 折线
			\draw[red] (0,3) -- (1,2) -- (2,3) -- (3,2) -- (4,2) -- (5,3);
			% 数据点的标签
			\node[below] at (1,0) {1};
			\node[below] at (2,0) {2};
			\node[below] at (3,0) {3};
			\node[below] at (4,0) {4};
			\node[below] at (5,0) {5};
			\node[left] at (0,1) {1};
			\node[left] at (0,2) {2};
			\node[left] at (0,3) {3};
			
		\end{tikzpicture}
	    \end{center}
            \caption{Diagram illustrating the change of $\varphi(3,i)$ with respect to i after i-th CNOT gates}
            \label{fig:7}
        \end{figure}
	
	By excluding cases where certain lines do not have a control qubit, we can find a counterexample that cannot achieve the desired permutation,  $\ket{1}\otimes\ket{1}\otimes\ket{0}$,This shows that this particular case is not valid.
	
	$\bullet(+1)+0+(-1)$
	
        \begin{figure}[H]
            \begin{center}
		\begin{tikzpicture}
			% 坐标轴
			\draw[->] (0,0) -- (5,0) node[right] {i};
			\draw[->] (0,0) -- (0,4) node[above] {$\varphi(3,i)$};
			
			% 数据点
			\draw[blue, fill=blue] (0,3) circle (2pt);
			\draw[blue, fill=blue] (1,2) circle (2pt);
			\draw[blue, fill=blue] (4,2) circle (2pt);
			\draw[blue, fill=blue] (5,3) circle (2pt);
			\draw[blue, fill=blue] (2,3) circle (2pt);
			\draw[blue, fill=blue] (3,3) circle (2pt);
			% 折线
			\draw[red] (0,3) -- (1,2) -- (2,3) -- (3,3) -- (4,2) -- (5,3);
			% 数据点的标签
			\node[below] at (1,0) {1};
			\node[below] at (2,0) {2};
			\node[below] at (3,0) {3};
			\node[below] at (4,0) {4};
			\node[below] at (5,0) {5};
			\node[left] at (0,1) {1};
			\node[left] at (0,2) {2};
			\node[left] at (0,3) {3};
			
		\end{tikzpicture}
	    \end{center}
            \caption{Diagram illustrating the change of $\varphi(3,i)$ with respect to i after i-th CNOT gates}
            \label{fig:8}
        \end{figure}
	
	Clearly there is no such a CNOT gate, because $\varphi(3,2)$=$\varphi(3,3)$=3. So the case is not valid.
	
	$\bullet(-1)+(+1)+0$
One can exclude this case by a similar argument to the previous cases. 

        \begin{figure}[H]
            \begin{center}
		\begin{tikzpicture}
			% 坐标轴
			\draw[->] (0,0) -- (5,0) node[right] {i};
			\draw[->] (0,0) -- (0,4) node[above] {$\varphi(3,i)$};
			
			% 数据点
			\draw[blue, fill=blue] (0,3) circle (2pt);
			\draw[blue, fill=blue] (1,2) circle (2pt);
			\draw[blue, fill=blue] (4,2) circle (2pt);
			\draw[blue, fill=blue] (5,3) circle (2pt);
			\draw[blue, fill=blue] (2,1) circle (2pt);
			\draw[blue, fill=blue] (3,2) circle (2pt);
			% 折线
			\draw[red] (0,3) -- (1,2) -- (2,1) -- (3,2) -- (4,2) -- (5,3);
			% 数据点的标签
			\node[below] at (1,0) {1};
			\node[below] at (2,0) {2};
			\node[below] at (3,0) {3};
			\node[below] at (4,0) {4};
			\node[below] at (5,0) {5};
			\node[left] at (0,1) {1};
			\node[left] at (0,2) {2};
			\node[left] at (0,3) {3};
		\end{tikzpicture}
	       \end{center}
            \caption{Diagram illustrating the change of $\varphi(3,i)$ with respect to i after i-th CNOT gates}
            \label{fig:9}
        \end{figure}

	\subsection{Six CNOT gates}
\label{sec:6cnot} 

    \quad In this subsection, within the 3-qubit system, we discuss the CNOT gates in Definition \autoref{the definition of CNOT}. According to the above description, we calculate the case of all one-five CNOT gates without a single gate, and none of them satisfy the condition. Therefore, six CNOT gates are necessary on this question. After calculation, we find all cases satisfying the CNOT gate in qubit three.

    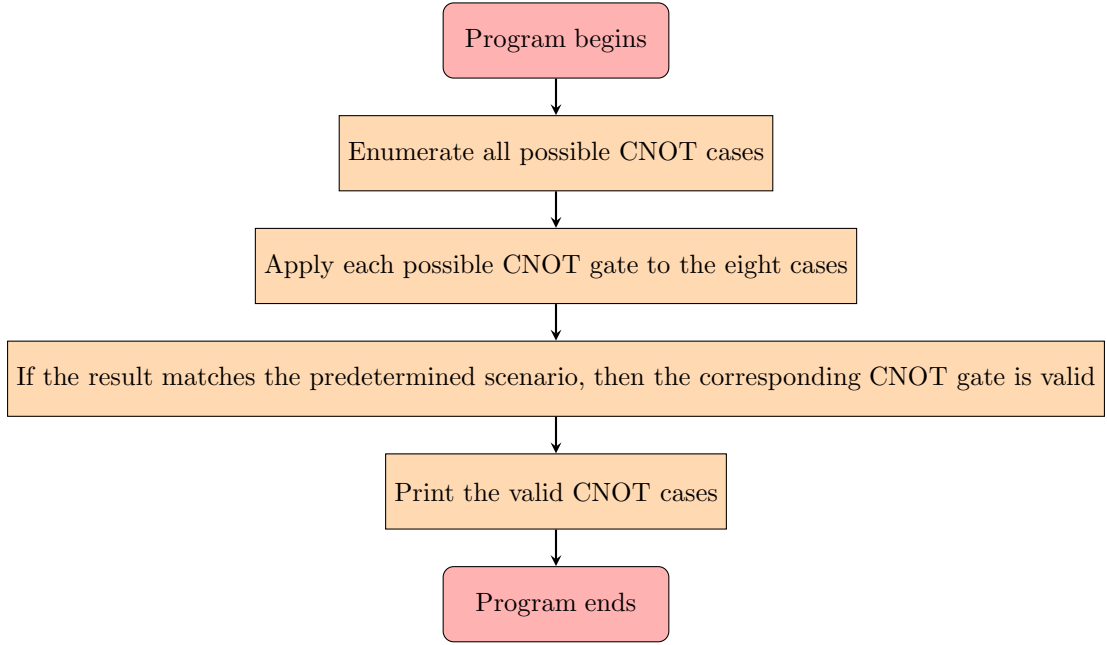
\begin{figure}[H]
    \centering
    \begin{tikzpicture}[node distance=1.5cm]
        % 定义流程图节点样式
        \tikzstyle{startstop} = [rectangle, rounded corners, minimum width=3cm, minimum height=1cm,text centered, draw=black, fill=red!30]
        \tikzstyle{process} = [rectangle, minimum width=2cm, minimum height=1cm, text centered, draw=black, fill=orange!30]
        \tikzstyle{arrow} = [thick,->,>=stealth]

        % 绘制流程图节点
        \node (start) [startstop] {Program begins};
        \node (process1) [process, below of=start] {Enumerate all possible CNOT cases};
        \node (process2) [process, below of=process1] {Apply each possible CNOT gate to the eight cases};
        \node (process3) [process, below of=process2] {If the result matches the predetermined scenario, then the corresponding CNOT gate is valid};
        \node (process4) [process, below of=process3] {Print the valid CNOT cases};
        \node (end) [startstop, below of=process4] {Program ends};

        % 绘制箭头
        \draw [arrow] (start) -- (process1);
        \draw [arrow] (process1) -- (process2);
        \draw [arrow] (process2) -- (process3);
        \draw [arrow] (process3) -- (process4);
        \draw [arrow] (process4) -- (end);
    \end{tikzpicture}
    \caption{Code flowchart}
    \label{figure:code flowchart}
\end{figure}

By traversing programmatically in \Cref{figure:code flowchart}, we filter out all cases of CNOT gate scenarios that satisfy the permutation condition, totaling 90 types. Next, we classify these 90 cases, by using the definition of CNOT gates and equivalence class in Definitions \autoref{the definition of CNOT} and \autoref{the definition of equivalence class}. For the CNOT gate scenarios that meet the requirements, if the first CNOT gate is A, then after one operation, the first CNOT gate becomes E, and after another operation, it becomes D. Therefore, the scenarios with the first gate being A, D, or E are equivalent, and the number of scenarios for these three cases is equal.
    
	Based on the above discussion on equivalence classes, these 90 scenarios can be divided into two equivalence classes: scenarios where the first CNOT gate is A, D, or E are equivalent, and scenarios where the first gate is B, C, or F are equivalent. Each equivalence class consists of fifteen scenarios. Consequently, these 90 scenarios are divided into completely distinct 30 equivalence classes. The enumeration of these cases is as follows.
	\begin{remark}
 \label{rmk:3qubit}
		We label each case in \Cref{tab:my_table} with the six gate numbers   provided from \Cref{fig:six CNOT gates in the 3-qubit system}, indicating them in sequential order from left to right. Simultaneously, we have provided three equivalent scenarios for one of these equivalence classes in \Cref{fig:equivalent class_example}, merging into equivalence classes‘AEFDCB’.
	\end{remark}

\begin{table}[htbp]
\label{tab:n=3for90}
    \centering
    \caption{The cases satisfying the condition for $n = 3$}
    \label{tab:my_table}
    
    \resizebox{\textwidth}{!}{%
    \begin{tabular}{cccccc} % 根据列数调整格式，c表示居中对齐，l表示左对齐，r表示右对齐
        \toprule % 表格顶部横线
         A & A & A & B & B & B  \\
        \midrule % 中间横线
         ABAEFE & ABFEFC & AEDFCB & BABEFE & BAFEFD & BFAEDF \\
         ABAFEF & ACBAFE & AEFDCB & BABFEF & BDABEF & BFAEFD \\
         ABEFCE & ACBFAE & AEFEDC & BAEBFE & BDAEBF & BFCEDA \\
         ABEFEC & AEBFCE & AFEDFC & BAEFED & BEFCED & BFECDA \\
         ABFAEF & AEBFEC & AFEFDC & BAFEDF & BEFECD & BFEFCD \\
                                          
        \bottomrule % 表格底部横线
    \end{tabular}
    }

\end{table}

\begin{theorem}
\label{thm:3qubit}
     Using six CNOT gates is both necessary and sufficient to achieve a prescribed permutation $(123)$ on a 3-qubit system. Furthermore, we have identified all 90 permutation diagrams and grouped them into the 30 equivalence classes in \Cref{tab:my_table}.
\end{theorem}

\section{Extension to the $n$-qubit case}
\label{sec:highdim}

In this section, we investigate the number $C(n)$ of CNOT gates required to implement the element in the $n$-qubit permutation group. This is done for the element $(123)$ in the last section, and an upper bound for $C(n)$ with $n\ge3$ has been shown in Lemma \autoref{le:C(n)le3(n-1)}. Because the number of elements in the $n$-qubit permutation group increases exponentially with $n$, it becomes difficult to respectively investigate the CNOT cost for the elements. For example, we need investigate the implementation of the element $(132)$. It can be directly shown that $(132)$ costs also exactly six CNOT gates by using a similar argument for proving Theorem \autoref{thm:3qubit}. Further, we shall claim in Theorem \autoref{thm:nqubit} that the so-called irreducible elements cost the same number of CNOT gates. Here the "irreducible" means that the element is not the union of two elements in smaller permutation groups. Otherwise, the element is reducible, e.g., $(12)(34)$ is a reducible element in the four-element permutation group.

To demonstrate our claim, we start with the case $n=4$. Evidently, there are $4!=24$ elements in the four-qubit permutation group. If the element is reducible, then one can show that the element is either $(ab)(cd)$ or $(abc)d$ where the distinct $a,b,c,d$ are $1,2,3,4$. It follows from Theorem \autoref{thm:3qubit} that such an element costs at most $3+3=6$ CNOT gates. So it suffices to investigate the number of irreducible elements in the four-element group. After enumeration, there are exactly six elements namely $(2341), (2413), (3142)$, $(3421), (4123), (4312)$.

\subsection{The implementation of multi-qubit swap gate}

%\red{(the following is revised by junchi and need be read).}

We apply the same method as in Section \ref{sec:three} to provide a proof for the case of four qubits. We put all the map of n CNOT gates in a set, named as $M_{n}$. Since CNOT gates are reflexive, any element of $M_{n}$ that adds two CNOT gates on the same qubits can be seemed as a case of $M_{n+2}$. Therefore, we can infer that $M_{n}$ is a subset of $M_{n+2}$ for any positive integer n. We speculate that 9 CNOT gates are needed in the case of 4 qubits, so we only need to explore the cases with 7 and 8 CNOT gates. If none of the cases of 7 and 8 CNOT gates can accomplish the goal, then any CNOT gates fewer than 9 cannot accomplish the goal.

%We can apply the same method as in Section \ref{sec:three} to address the case of 4 qubits, and 

Furthermore, we provide an alternative approach to solving this problem in the following sections, leading to our conclusion. For (1234), it has been demonstrated that 9 CNOT gates are needed. We name (1234) clearly as 1234$\xrightarrow{}$2341. Meanwhile,we can switch qubit 1 and 2. We get 2134$\xrightarrow{}$3241, which does not alter the required number of CNOT gates. Moreover, we can switch qubit 2 as qubit 1, qubit 1 as qubit 2, and we get 1234$\xrightarrow{}$3142, which is (3142). We simply denote such an operation as $\eta_{12} = 3142$, so (3142) requires at least 9 CNOT gates. In a similar way, $\eta_{13} = 4123, \eta_{14} = 4312, \eta_{23} = 3421, \eta_{24} = 4123, \eta_{34} = 2413$. As for n = 4, all the irreducible cases are as follows: $(2341), (2413), (3142)$, $(3421), (4123), (4312)$. Reducible cases can be accomplished within 6 CNOT gates. In summary, $C(4) = 9$.

However, when discussing the case of four qubits, we find that as the number of qubits increases, the complexity of the problem rises sharply. For example, when considering scenarios involving 7 and 8 CNOT gates, we have to discuss too many cases, so we hope to find a more expedient approach. Different from the approach taken in Section 3, which focuses on verifying the 32 vectors within a specific circuit configuration. We define the quantum information at the input as abcde and examine the minimum number of CNOT gates required to achieve the output configuration bcdea.

Firstly, the sequence abcde can be represented in binary as 00001, 00010, 00100, 01000, and 10000, respectively. This assignment is designed to maintain the integrity of each qubit during the CNOT operations, which are essentially XOR operations. This means that one bit will not transform into another, preserving the distinct identity of each qubit. This binary representation allows us to establish a direct correspondence between an integer and the quantum state. For instance, a state represented by 13 in decimal (which equals $2^{3} + 2^{2} + 2^{0}$) corresponds to the quantum state comprising qubits a, c, and d.Next, we consider each CNOT gate operation iteratively. To determine the minimum number of CNOT gates required, we simulate each CNOT gate operation iteratively. This process continues until the current quantum state matches the desired final state. During each iteration, we record the number of CNOT gates used. By implementing this process in code, we efficiently track the transformations of the quantum state. For the specific case of transforming abcde to bcdea, we find that the minimum number of CNOT gates required is 12.

In the following section, we explore whether the minimum number of CNOT gates required in different configurations is also 12. Through a series of equivalence discussions in the subsequent chapter and further simulations, we aim to prove that the function $C(5)$, representing the maximum number of CNOT gates needed for any transformation in a 5-qubit system, equals 12.

\subsection{The implementation of n-qubit swap gate}

   \quad Using swap circuits and renaming techniques, we consider the general $n$ qubit case. First, we define reducible cases as follows.

 \begin{definition}
    \label{df:multiqubit}
        If some kinds of reducible arrangement of transformation of case $(12...n)$ can be divided into the product of two cases $((12...k) and (k+1...n))$ of conversion, then for this transformation there is
\begin{equation}
C(n)=C(k)+C(n-k)\leq3(k-1)+3(n-k-1)=3(n-2).        \end{equation}
\end{definition}

    The following studies are all about irreducible cases, such as $(12...n)$. It can be regarded as connecting the n points of positive n equally divided on the circle, and the connection is relatively connected, otherwise the transformation is reducible, as shown in the figure below.
    
    \Cref{3-4-1} illustrates the case $(12...n)$. Apparently it is irreducible.
    
    % \begin{figure}[H]
        
    %    \centering
    %    \includegraphics[width=10cm]{3-4-1}
    %    \caption{n points on the circle}
    %    \label{3-4-1}
    %\end{figure}  
    \begin{figure}[H]
        \centering
        \begin{tikzpicture}[scale=1.5]
            \foreach \y in {0,1,2} 
            {
                
                \draw (0,0) circle [radius=2.5cm];
                \fill (-2.5,0) circle(2pt);
                \fill (1.5,2) circle(2pt);
                \fill (-2,1.5) circle(2pt);
                \fill (1.5,-2) circle(2pt);
                
                \node at (-2.8,0){1};
                \node at (1.7,2.2){3};
                \node at (-2.2,1.7){2};
                \node at (1.7,-2.2){n};
                \draw (-2.5,0) -- (-2,1.5);
                \draw (-2,1.5) -- (1.5,2);
                \draw (1.5,-2) -- (-2.5,0);

            }		
        \end{tikzpicture}	
        \caption{n points on the circle}
        \label{3-4-1}
    \end{figure}
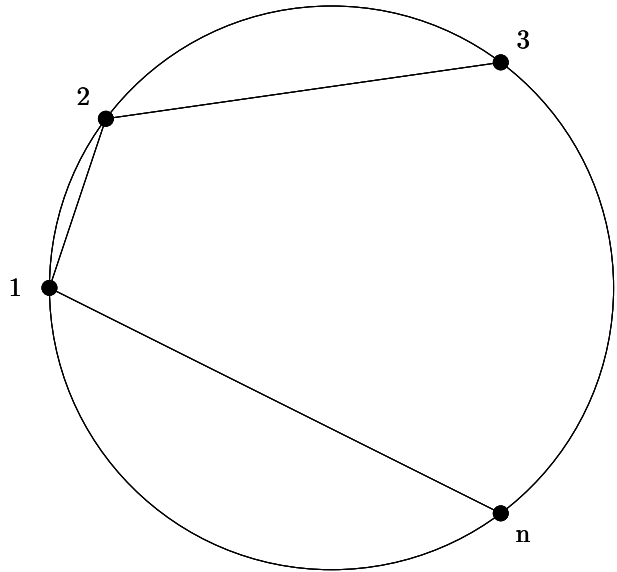

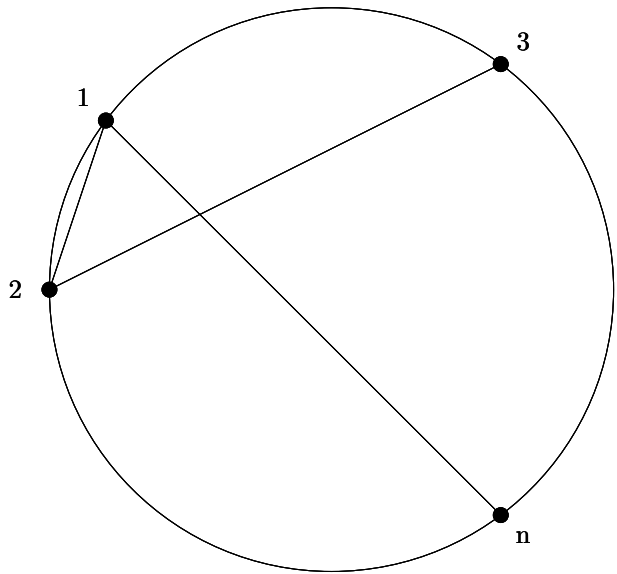
\begin{figure}[H]
        \centering
        \begin{tikzpicture}[scale=1.5]
            \foreach \y in {0,1,2} 
            {
                
                \draw (0,0) circle [radius=2.5cm];
                \fill (-2.5,0) circle(2pt);
                \fill (1.5,2) circle(2pt);
                \fill (-2,1.5) circle(2pt);
                \fill (1.5,-2) circle(2pt);
                
                \node at (-2.8,0){2};
                \node at (1.7,2.2){3};
                \node at (-2.2,1.7){1};
                \node at (1.7,-2.2){n};
                \draw (-2.5,0) -- (1.5,2);
                \draw (-2.5,0) -- (-2,1.5);
                
                \draw (1.5,-2) -- (-2,1.5);

            }		
        \end{tikzpicture}	
        \caption{The connecting line of n points after exchanging positions}
        \label{3-4-2}
    \end{figure}

    Obviously, the number of such a transformation is $(n - 1)!$. We leave the line segment relationship in \Cref{3-4-1} unchanged and transform the position relationship between 1 and 2 to get \Cref{3-4-2}.

    %\begin{figure}[H]
    %    \centering
    %    \includegraphics[width=10cm]{3-4-2}
    %    \caption{The connecting line of n points after exchanging positions}
    %    \label{3-4-2}
    %\end{figure}
    
\begin{figure}[H]
        \centering
        \begin{tikzpicture}[scale=1.5]
            \foreach \y in {0,1,2} 
            {
                
                \draw (0,0) circle [radius=2.5cm];
                \fill (-2.5,0) circle(2pt);
                \fill (1.5,2) circle(2pt);
                \fill (-2,1.5) circle(2pt);
                \fill (1.5,-2) circle(2pt);
                
                \node at (-2.8,0){1};
                \node at (1.7,2.2){3};
                \node at (-2.2,1.7){2};
                \node at (1.7,-2.2){n};
                \draw (-2.5,0) -- (1.5,2);
                \draw (-2.5,0) -- (-2,1.5);
                
                \draw (1.5,-2) -- (-2,1.5);

            }		
        \end{tikzpicture}	
        \caption{Renamed n-point concatenation}
        \label{3-4-3}
    \end{figure}
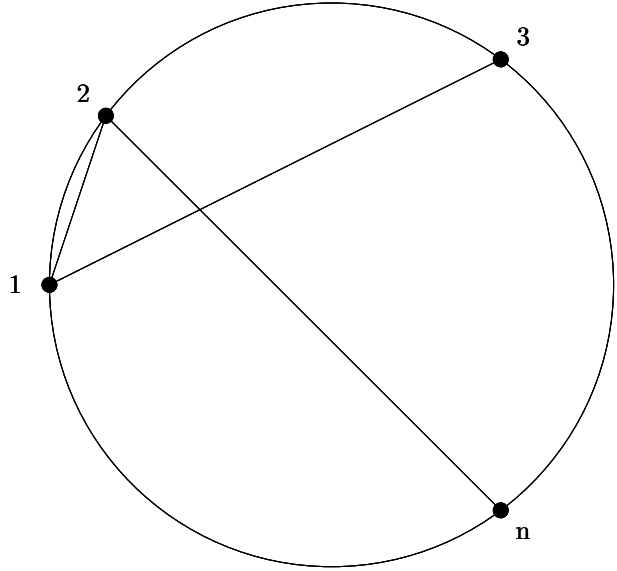

    Then we rename 1 and 2, so we get \Cref{3-4-3}. In summary, by employing swapping circuits and renaming techniques, $(123...n)$ transforms into $(213...n)$.

    %\begin{figure}[H]
    %    \centering
    %    \includegraphics[width=10cm]{3-4-3}
    %    \caption{Renamed n-point concatenation}
    %    \label{3-4-3}
    %\end{figure}

    The resulting $(213...n)$ in \Cref{3-4-3} is still connected and irreducible, and this operation is remembered as one operation. It's obvious to know that any two points which lined up can operated adjacent within limited operations(at most n-2 operations). Thus, we can infer that for any kind of irreducible and connected transformation of the figure above, it is always possible to perform at most finite operations to transform the set of graphs into the situation as shown in \Cref{3-4-1}.

    According to the definition, we can assume that, the $C(n)$ of completing the operation of $(12...n)$ is $3(n-1)$. As a consequence, we can conclude, the irreducible transformation $C(n) = 3(n-1)$. Then, the remaining reducible C(n) is less than $3(n-2)$, i.e $C(n)\leq 3(n-2)$. As a result, $C_{max}=3(n-1)$.
    
    According to the above analysis, we find that the name of the point on the circle has no influence on the whole circle, that is, there is equivalence between n-qubit cases.
    \begin{theorem}
\label{thm:nqubit}
In the $n$-qubit permutation group, 
    
(i) the implementation of irreducible elements costs the same number of CNOT gates upper bounded by $3n-3$,

(ii) the implementation of reducible elements costs the number of CNOT gates upper bounded by $3n-6$. 

\end{theorem}
Theorem \autoref{thm:nqubit} (i) shows that, it suffices to investigate the number of CNOT gates for the element $(12...n)$ in the $n$-element permutation group.

\section{Conclusion}
\label{sec:con}
We have shown that six CNOT gates are both necessary and sufficient for the implementation of permutation group of three elements. We also have provided all circuits realizing the element by using exactly six CNOT gates. Our results have been extended to multiqubit system. A problem arising from this paper is to implement or show the non-existence of the three-qubit swap gate using local unitary gates and less than six CNOT gates. Another interesting problem is whether the CNOT gate can be replaced by another gate which may cost fewer than that of CNOT gates in a quantum circuit. 
	
\section*{Acknowledgements} 

The paper was supported by the NNSF of China (Grant No. 11871089). 

\bibliographystyle{unsrt}
		\bibliography{km}

\end{document}